\documentstyle[aps,floats,prl,graphicx]{revtex}
\def\ts{{\bf t}(s)}
\def\t{{\bf t}}
\def\l{\ell}
\def\ls{{\ell^*}}
\def\lss{\ell^{**}}

\begin{document}
\title{Writhing Photons and Berry Phases in Diffusive Wave Scattering}
\author{A.C.~Maggs, V.~Rossetto\\
\it Laboratoire de physico-chimie th\'eorique, ESA-CNRS 7083\\
\it ESPCI, 10 Rue  Vauquelin, Paris, 75005, France.}

\twocolumn[\hsize\textwidth\columnwidth\hsize\csname @twocolumnfalse\endcsname

\maketitle

\begin{abstract}
  We study theoretically the polarization state of light in multiple
  scattering media in the limit of weak gradients in refractive index.
  Linearly polarized photons are randomly rotated due to the Berry
  phase associated with the scattering path. For circularly polarized
  light independent speckle patterns are found for the two helical
  states.  The statistics of the geometric phase is related to the
  writhe distribution of semiflexible polymers such as DNA.
\end{abstract} 

\vskip 1cm
]
\bigskip

Multiple light scattering techniques \cite{etienne1} allow the
experimentalist to probe deep into opaque, strongly scattering samples
giving information on both static and dynamic correlations. The
preferred way of interpreting these experiments is via a field of {\sl
  scalar}\/ photons \cite{weitz} which satisfy, in the simplest
analytic treatments, a diffusion or transport equation. The phase of
the wave associated with each path is just proportional to the path
length sampled by each photon.  In this letter we show the importance
of geometric phases in the evolution of photons in multiple scattering
which have as their origin the vector nature of light. We give
explicit expressions for the probability distribution of this extra
phase in transmission geometry.  Rather surprisingly, whilst the basic
propagation laws for light in inhomogeneous media go back over 60
years to the work of Rytov \cite{r1,berry2}, the full implication of
his results has not yet been exploited in diffusive wave scattering.
In this letter, we consider the consequences of Rytov's observations
on the scattering of circularly and linearly polarized light.

We first summarize our main results before giving a more extended
discussion: For {\sl circularly polarized light}\/ the photon
helicity is a conserved quantum number in systems with weak
inhomogeneities \cite{r1} within the eikonal approximation. Each
possible, multiple scattered, path through a sample is highly tortuous
and thus has associated with it a writhe $\phi_i/2\pi$
\cite{berry2,berry}. A Berry phase, $e_i \phi_i$, then adds to the
simple geometric optical path $\psi_i=q \ell_i$, calculated in a
scalar theory, where $q$ is a wavenumber, $\ell_i$ the optical path
length and $e_i=\pm 1$ is the photon helicity.  In a transmission
geometry this leads to two possible speckle patterns for a given
arrangement of scatters as a function of the polarization state of
the light.  For {\sl linearly polarized light}\/ the situation is more
subtle: For each individual path the geometric phase rotates the plane
of polarization \cite{optics}.  However the final experimental
measurement is a measure of intensity, summing over all possible
paths. This sum leads to a final polarization state which is in
general elliptical rather than linear.  We shall thus characterize the
evolution of a linearly polarized state as a distribution of polarization
states on the Poincar\' e sphere.  Note that a close analog of Rytov's
result has been studied in the propagation of polarized light along a
fiber optic: A beam of light transmitted along a fiber \cite{optics}
acquires a non-trivial phase when the tangent of the fiber ${\ts}$
plotted on a unit sphere encloses a non-zero area. This experiment is
in many ways the simplest realization of the Berry phase and is clearly
analogous to the propagation of a single ray in a weakly scattering
medium.

We note that there is an important {\sl qualitative} difference
between the evolution of circular and linear polarization states in
weakly scattering media. For the former the helicity can be preserved
over arbitrarily large distances, whereas we shall conclude that the
Berry phase associated with linearly polarized light always leads to a
state of random polarization.  We thus suggest that the natural
setting for experimentally studying polarization effects in multiple
scattering media is in critical, opalescent samples where the gradual
density gradients allow an exact formulation of the polarization
statistics without backscattering from interfaces.

Technically we treat the problem of summing over photon paths via a
mapping onto a semiflexible polymer, which treats multiple weak
scattering of photons as an {\sl angular} diffusion process. The Berry
phase is calculated from the writhe \cite{fuller} of the photon path
using methods introduced to study the statistical mechanics of DNA and
other stiff molecules. Our considerations also link up with remarks of
\cite{mackintosh} who gave a local argument for the evolution of the
polarization vector, in the context of multiple scattering, equivalent
to that of Berry's.

Since the DWS technique is very often used to study the properties of
colloidal systems we start by explaining how some of the above ideas
are applied to such multi-center scattering systems, where the mapping
onto a torsionally rigid semiflexible polymer is particularly direct
and simple to understand. In the theory of DWS from colloidal samples
one defines two characteristic distances. The first, $\ell$, is the
distance between two collisions between a photon and a scattering
center. The second, $\ls$, measures the distance over which a photon
must travel in order to forget its initial direction of propagation.
In a simple analogy with stiff polymers one can consider that the
length $\l$ corresponds to a monomer size while $\ls$ is equivalent to
the persistence length of the polymer.  In strongly scattering media,
with high contrast between the refractive indexes between inclusions
and background these two lengths are comparable.  However by using
particles large compared with the wavelength of light and with low
contrast between the dielectric properties of the two media we can
easily find samples for which $\ls/\l$ is of order ten.  In recent
experiments \cite{etienne2} in large droplet helium aerosols it is has
proven possible to increase $\ell^*/\ell$ to over $200$.  Motivated by
this last experimental system we present here arguments as to the
polarization statistics of light scattered in a regime of intermediate
sample thickness, $L$, such that $\ell \ll L \leq \ell^*$ in
transmission geometry. In such samples photons are scattered many
times, but are still propagating largely in the forward direction.
This is exactly the geometry studied experimentally in
\cite{etienne2}.

We now derive the mapping which allows us to transpose results known
from the statistical mechanics of stiff polymers, including now the
torsion as well as bending degrees of freedom.  Following
\cite{etienne2} we shall model the medium as an ensemble of randomly
oriented interfaces neglecting the spherical structure of the aerosol.
This was shown to give a good qualitative description of the
scattering statistics. Consider the collision between a photon and a
single interface. The photon can either be reflected, and thus
deviated by a large angle, or refracted by a small angle which can be
calculated using geometric optics. The probability of reflection, for
a typical impact parameter, is comparable to
\begin{math}
  {\cal R} = (n-1)^2/(n+1)^2
\end{math}
where $n$ is the refractive index of the inclusions relative to that
of the background medium. If we write
\begin{math}
  \epsilon = (n-1)
\end{math}
then we see that the intensity of the direct beam should decay over a
length which is at most
\begin{math}
  \ell_{ref}=\ell/\epsilon^2
\end{math}
due to back scattering. A background of photons scattered through
large angles is indeed observed experimentally, in addition to the
main forward beam \cite{etienne2}.

Most of the beam is refracted at the interface. For typical impact
parameters the angle of deviation should be comparable to $\epsilon$.
Since successive deviations of a photon are independent this implies
that the direction of propagation of the photon diffuses as it
propagates into the medium with angular diffusion coefficient $D \sim
\epsilon^2/\ell$.  Thus photons which are not directly reflected at a
surface also turn over a length comparable to $\ell/ \epsilon^2$ which
is thus our estimate of $\ls$ in this weakly scattering limit.

Let us now consider the evolution of an incident linearly polarized
beam.  The reflectivity of an interface is a function of the plane of
polarization with respect to the surface; the transmitted beam has a
modified polarization state. In the hypothetical case of perfect
transmission the plane of polarization of the refracted beam evolves
by parallel transport \cite{mackintosh,acm}. Due to the
reflections the transmitted amplitudes of the two polarization
components (defined relative to the local surface orientation) are
comparable to $1- O(\epsilon^2)$. This leads to a rotation of plane of
polarization of the transmitted light by an angle $\pm O(\epsilon^2)$
compared to a parallel transported state.  In the analogy with a stiff
polymer this corresponds to excitation of a {\sl torsional}\/ mode. In
stiff polymers one can define two independent persistence lengths for
bend and torsional degrees of freedom, $l_p$ and $l_t$.  These lengths
are usually comparable. For the case of multiple light scattering we
see that $\l_p \sim \ls \sim \ell/\epsilon^2$ whereas $\l_t \sim
\ell/\epsilon^4$. The ``torsional'' degree of freedom for photons is
frozen out at low dielectric contrast as $l_t \gg l_p$, so that as
$\epsilon \rightarrow 0$ parallel transport of the transmitted
component becomes exact, in agreement with Rytov's treatment of
Maxwell's equations in the eikonal approximation.  We have thus
extended the analogy between stiff polymers \cite{nelson} and
diffusing photon paths by studying the torsional degrees of freedom of
the polymer, which are entirely analogous to the polarization states
of photons.

The tangent to the curve ${\bf t}(s)$ is a unit vector and can be
considered as living on a unit sphere. We now use the Berry formula
linking the area enclosed by the curve ${\bf t}(s)$ on this sphere
with the geometric phase to calculate the probability distribution of
phases associated with a samples in the limit $\ell \ll L \ll \ls$.
The orientation of a photon diffuses \cite{acm} as a function of
curvilinear distance, so that the probability of orientation obeys the
equation
\begin{equation}
{\partial P(\t ,s) \over \partial s}
=
{D\over 2 }
\nabla^2P(\t ,s) \ ,
\label{fp}
\end{equation}
with $\nabla^2$ the spherical Laplacian and $s$ the optical path
length.  When the sample is thin compared with $\ls$ the photons do
not diffuse very far from their original propagation direction.  We
can thus calculate the phase contribution from the Berry phase by
looking at the problem of diffusion on a local, planar approximation
to the sphere. The probability distribution for the area of random
loops on a plane was solved by Levy \cite{acm,levy} with the result
for the probability distribution for the Berry phase:
\begin{equation}
{\cal P} (\phi) =  
 {1 \over 2DL \cosh^2( \phi /DL )}
\quad .
\label{writhedis}
\end{equation}
From this expression we find the mean square phase as
\begin{equation}
\langle \phi^2\rangle = {  \pi^2 L^2D^2 \over 12 } \ .  
\label{shortwrithe}
\end{equation}

We proceed by studying the Jones vectors describing the electric field
of a coherent light source.  Vertically and horizontally polarized
light is described by the vectors $j_v=(1,0)$ and $j_h=(0,1)$
respectively, whereas circular light corresponds to the vectors
$j_{\pm}=1/\sqrt{2}\;(1, \pm i)$. We are ultimately interested in the
intensities of the various polarization states of the transmitted beam
which are most easily visualized via the Poincar\' e sphere, fig (1).
The three axes correspond to a series of measurements $i_1=
I_0-I_{90}$, $i_2=I_{45} -I_{-45}$ and $i_3=I_{+} -I_{-}$. Here,
$I_{0,\pm 45,90}$ are the normalized intensities measured with with a
linear polarizer inclined at the subscripted angle and $I_{\pm}$ is the
intensity measured with circular analyzers.

As stated above for {\sl circularly polarized light}\/ $e_i \phi_i$
can simply be added to the phase $\psi_i$ so that the transmitted
Jones vector can be written in the form
\begin{equation}
e^{i( \psi_i \pm \phi_i)}  
 {1 \over{\sqrt{2}}}
\left(
\begin{array}{c}
  1 \\
  \pm i
\end{array} \right)
\end{equation}
for the incident state $j_{\pm}$. We now sum over all possible paths
to find the total scattered amplitude: The sum of the individual
circularly polarized states also gives rise to circularly polarized
light as the final state.  When $L/l* \sim 2$ the geometric phase
introduces a relative phase $2 \phi_i$ between the right and left
states comparable to $\pi$.  We thus understand that the speckle
pattern for the two helical states is similar for very thin samples,
whereas for thicker samples we find two independent intensity
distributions for each circularly polarized state.
\begin{figure}
\begin{center}
  \includegraphics[scale=0.2]{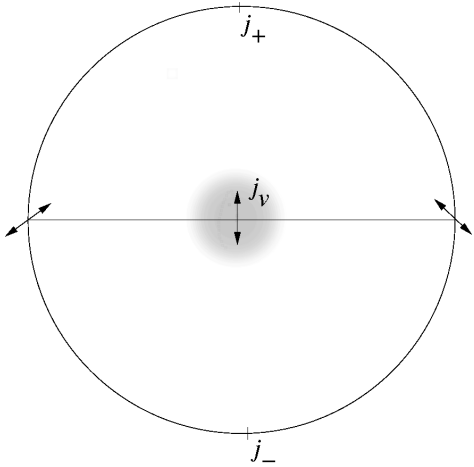}
\caption{ 
  The Poincar\' e sphere describing the polarization state of the
  light.  Linearly polarized states correspond to points along the
  equator while the north and south poles, labeled $j_+$ and $j_-$
  correspond to circularly polarized light.  Other states are
  elliptically polarized.  Multiple scattering of the initial state
  $j_v$ describing linearly polarized light leads to a circularly
  symmetric distribution of polarization states centered on $j_v$ in
  samples where $L/l*<1$}
\label{fig}
\end{center}
\end{figure}

If we illuminate with {\sl linearly polarized} light $j_v=(1,0)$, each
transmitted photon is described by its writhe $\phi_i$, and the total
phase $\psi_i$ so that the transmitted state is
\begin{equation}
 e^{i \psi_i} \left(
\begin{array}{c}
  \cos{\phi_i} \\
  \sin{\phi_i}
\end{array} \right) \approx
e^{i \psi_i} \left(
\begin{array}{c}
  1 \\
  \phi_i
\end{array} \right)
\label{amplitude}
\end{equation}
where we have specialized to samples with $L/\ls <1$.  $\psi$ is
calculated from the statistics of longitudinal fluctuation of a stiff
polymer by writing the path length as
\begin{math}
  \Delta = \int \; ds \sqrt{1+ (d {\bf r}_{\perp}/ds)^2}.
\end{math} with ${\bf r}_{\perp}$ the deviation of the path from
a straight line.  By expanding the square root and studying the
correlation function $\langle \Delta^2 \rangle$, \cite{gittes} one
finds that the fluctuations of path length, $\ell_i$, for crossing a
sample of thickness $L$ are of order $L^2/\ls$, when $L<\ls$.  Thus
$\psi_i$ is very large, so that $\psi_i \; {\rm mod}(2\pi)$ is very
nearly uniformly distributed.

If we sum the amplitude eq.  (\ref{amplitude}) over a large number of
independent paths we see that we find a vector of the form
\begin{equation}
 j_f=A e^{i \psi_0} \left(
\begin{array}{c}
  1 \\
  \phi_0 e^{i \psi_1}
\end{array} \right)
\end{equation}
A relative phase between the two components of the vector $j_f$
develops because the random sign of $\phi_i$.  For small, fixed
$\phi_0$ the vector $j_f$ describes a circle of radius $2\phi_0$ on
the Poincar\' e sphere as $\psi_1$ varies between $0$ and $2\pi$.  We
note, in passing, that the full joint distribution function of the
writhe and path length, ${\cal P}(\phi,\psi)$, is closely linked with
the force-torsion response curves measured in DNA micromanipulation
experiments \cite{nelson}

Thus $A$ and $A \phi_0$ are random variables with $\phi_0 \sim
\sqrt{\langle \phi^2 \rangle} \sim L/\ls$, eq.  (\ref{shortwrithe}).
$\psi_1$ is again uniformly distrubuted.  For thin samples this random
Jones vector is distributed in a circular disk, centered on the
initial vector $j_v$, fig(1). The combination of a Berry phase
combined with the widely distributed optical path leads to a state of
elliptical polarization.  For thick samples we find a (non-uniform)
distribution of vectors over the whole Poincar\' e sphere leading to
loss of memory of the initial polarization state in samples thicker
than a few times $\ls$.  Detailed simulations \cite{these,maynard}
have been performed to study the decay of polarization in
colloidal systems in transmission geometry with parameters
corresponding to polystyrene beads in water. It was indeed observed
numerically that the evolution of the polarization state proceeds by
the formation of a circular distribution on the Poincar\' e sphere.

The simulations in \cite{these,maynard} also found that the
polarization state of circularly polarized light decays in thick
samples.  The authors defined new lengths $\lss_{circ}$ and
$\lss_{plane}$, while noting that $\lss_{circ} > \lss_{plane}$, over
which both circular and linear polarization decay. We understand the
result that $\lss \sim \ls$ even for circularly polarized light as
being due to back reflection at surfaces. As argued above this leads
to a decay in intensity (and helicity flipping) of the beam over the
length $\ell_{ref}$ comparable $\ls$. If we were to perform similar
experiments on samples with slow variations in refractive index, for
instance critical, opalescent samples we expect that the reflection
occurring at interfaces should be suppressed due to critical
broadening. In this case the helicity of photons is preserved over a
length which is presumably exponentially large in the eikonal
expansion parameter \cite{r1}.  Thus $\lss_{circ}/ \ls$ can be
arbitrarily large. We reiterate that for experiments with linearly
polarized light the geometric phase, together with the phase $\psi_1$
due to the varying path lengths implies that memory of the linear
state must always be lost on a length scale comparable to $\ls$.

To conclude we have reached an understanding as to the evolution of
the polarization states in multiple scattering situations via a
mapping onto a writhing polymer. Despite the extreme simplification of
our arguments the results are in agreement with detailed simulations
based on a partial wave simulation using Mie diffraction theory
\cite{these,maynard}.  Our results are valid for {\sl coherent} light
sources; the essential step was the combination of the {\sl
  amplitudes} of the Jones vectors. In scattering with {\sl
  incoherent} sources it is rather the Stokes parameters which should
be combined, presumably leading to multicomponent transfer theories
such as those discussed in \cite{gorod}. The contrast between multiple
polarized scattering with coherent and incoherent sources deserves
closer study.

In our rather crude description of the colloidal regime we have
neglected diffraction effects, limiting our treatment to extremely
large droplets, however many of the qualitative conclusions should
hold even for more moderate particle sizes where diffraction is more
important: similar conclusions as to the statistics of the photon
paths were deduced in \cite{mackintosh} using the Born scattering
approximation.  It would be particularly interesting to perform
simulations in the limit of low optical contrast in order to bring out
some of the scaling regimes which may exist in this limit.

Finally we have not considered the nature of the non Gaussian
statistics of the polarization states due to the caustics presumably
present in weak scattering limits discussed in this letter.  A final
question that we leave open is the nature of two time correlation
functions of the polarization state: As scattering centers move both
the writhe of the photons paths and the phase associated with each
path vary.  Does this variation contain any interesting information on
the dynamics of colloidal systems?

\vskip 0.5cm We thank Bart van Tiggelen for introducing us to
\cite{these} together C. Derec and P.-E. Wolf for discussions on light
scattering methods.

\end{document}